 \documentclass[12pt,preprint]{aastex}

\shorttitle{Emissions from FGRB}
\shortauthors{Xu et al.}

\begin{document}
\title{Failed Gamma-Ray Bursts: Thermal UV/Soft X-ray Emission Accompanied by Peculiar Afterglows}
\author{M. Xu\altaffilmark{1,2,3}, S. Nagataki\altaffilmark{2}, Y. F. Huang\altaffilmark{1}, and S.-H. Lee\altaffilmark{2}}
\altaffiltext{1}{Department of Astronomy, Nanjing University, Nanjing 210093, China; hyf@nju.edu.cn}
\altaffiltext{2}{Yukawa Institute for Theoretical Physics, Oiwake-cho, Kitashirakawa,
Sakyo-ku, Kyoto 606-8502, Japan}
\altaffiltext{3}{Department of Physics, Yunnan University, Kunming 650091, China}


\begin{abstract}
We show that the photospheres of ``failed'' Gamma-Ray Bursts (GRBs),
whose bulk Lorentz factors are much lower than 100, can
be outside of internal shocks. The resulting radiation from the
photospheres is thermal and bright in UV/Soft X-ray band. The
photospheric emission lasts for about one thousand seconds
with luminosity about several times $10^{46}$ erg/s. These events can
be observed by current and future satellites. It is also shown that
the afterglows of failed GRBs are peculiar at the early
stage, which makes it possible to distinguish failed GRBs from
ordinary GRBs and beaming-induced orphan afterglows.
\end{abstract}


\section {Intronduction}

Gamma-Ray Bursts (GRBs) are the most powerful explosion in
the universe. The origin of prompt emission remains
unresolved, owing to the fact that the prompt emission has a large
explosion energy showing non-thermal spectrum with rapid
time-variabilities.

It is widely accepted that the prompt emission is
coming from a highly-relativistic flow, because it can reduce the optical depth
of the flow which makes the radiation spectrum non-thermal (Rees \& M\'{e}sz\'{a}ros 1994).
In fact, in the internal shock scenario, which is one of the most promising
scenarios, the relativistic shells collide with each other after the
system becomes optically thin (e.g. Piran 1994 for a review).
However, it was pointed out by M\'{e}sz\'{a}ros \& Rees (2000) that even such a
relativistic flow
should have a photosphere inevitably and thermal radiation should
be coming from there. Since then, there have been many
theoretical (Daigne \& Mochkovitch 2002; Pe$'$er et al. 2006; Pe$'$er 2008; Pe$'$er \& Ryde 2011)
and observational (Ghirlanda et al. 2003; Ryde 2004, 2005; Ryde et al. 2010;
Guiriec et al. 2010; Ryde et al. 2011)
studies on how the thermal component contributes
to the prompt emission (or the precursor).
Nowadays, the internal shock model with a photosphere is frequently
discussed (e.g. Toma et al. 2011; Wu \& Zhang 2011).
In this picture, the radius of the photosphere, $R_{\rm PS}$, is usually
smaller than the radius $R_{\rm IS}$ where internal shocks are
happening.

The above scenario is based on the assumption that the bulk Lorentz
factor of the jet is as large as 100-1000.
But what happens if the bulk Lorentz factor is not so high?
Theoretically, it is natural to consider such a case, because it is
very hard to realize such a clean, highly-relativistic flow. Especially, in
case of long GRBs, some bursts are at least coming from the death of massive
stars where a lot of baryons should be surrounding the central
engine (MacFadyen \& Woosley 1999; Proga et al. 2003; Nagataki et al. 2007; Nagataki 2009, 2010).
Thus we can expect there are a lot of ``failed GRBs''
that have dirty, not so highly-relativistic flows in the
universe (e.g. Dermer et al. 1999; Huang et al. 2002; Paragi et al. 2010; Xu et al. 2011).
Recently, Bromberg et al. (2011a, 2011b) suggested the existence of a large population of
failed GRBs if the jets failed to break out of the progenitor stars in collapsar model.

\begin{figure}
\includegraphics[scale=0.8]{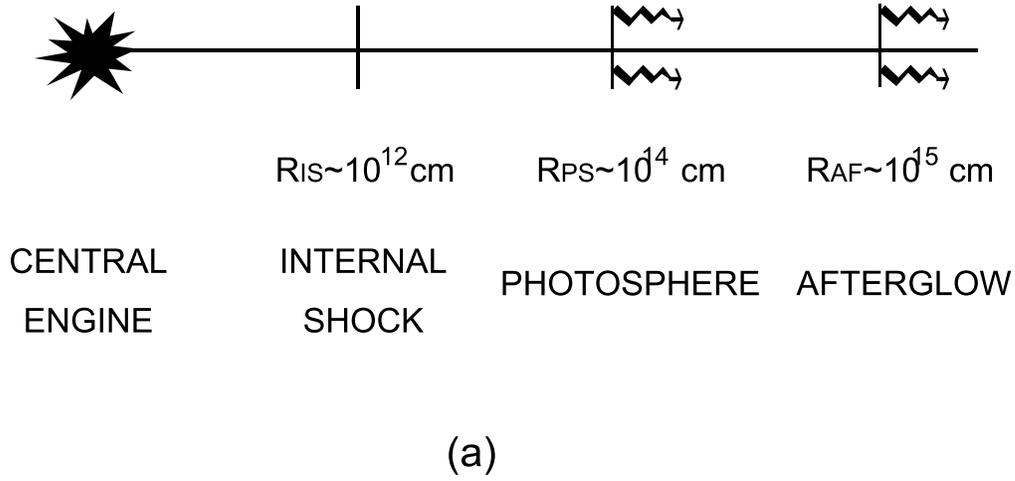}
\hspace*{-0.8cm}\includegraphics[scale=0.80]{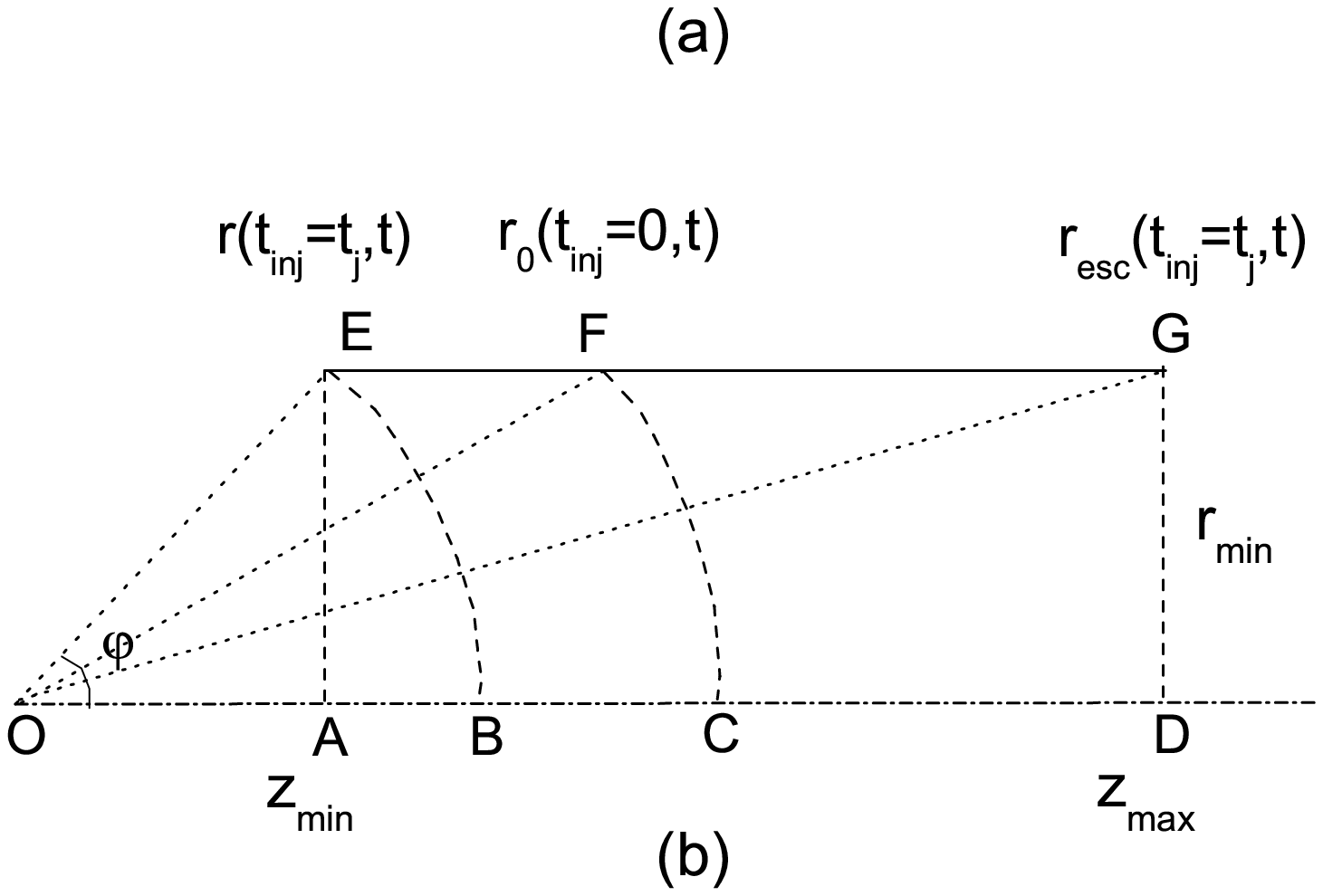}
\caption{\label{fig:cartoon}
(a) A sketch illustrating the emission regions of a failed-GRB.
Note that the internal shock radius is smaller than the photosphere radius.
(b) Schematic diagram showing the path of a photospheric photon escaping
from the ejecta in the stellar frame. A photon emitted at point E will
escape from the ejecta at point G.}
\end{figure}

Qualitatively, $R_{\rm IS}$ becomes smaller if the bulk Lorentz factor
of the flow is smaller, while $R_{\rm PS}$ increases with the decreasing of bulk Lorentz factor.
Thus we can expect that the photosphere will become outside of
the internal shock region for some lower Lorentz factors
(see Fig. \ref{fig:cartoon}a). In such a
case, $\gamma$-rays from the internal shocks cannot escape. Instead,
softer thermal radiation from the photosphere
followed by an afterglow will be seen.

We note that such a situation was considered in case of a
``successful GRB'' (Rees \& M\'{e}sz\'{a}ros 2005; Lazzati et al. 2009;
Mizuta et al. 2011; Nagakura et al. 2011; Ryde \& Pe$'$er 2009; Ryde et al. 2010, 2011).
In the photospheric model, it is
considered that the photospheric emission itself is the origin of the prompt
emission: its spectrum is modified to non-thermal due to the
heating by relativistic electrons that are produced at the internal
shocks inside the photosphere (Beloborodov 2010; Vurm et al. 2011).
Thus the photospheric model can
explain normal successful GRBs.

Here, in this study, we consider the case of ``failed GRBs''.
We show that the direct emission from the photosphere of a failed-GRB
will be a UV/soft X-ray burst, followed by
an afterglow with a peculiar
spectrum at early stage. It is also shown that the afterglows of failed GRBs can
be distinguished from ordinary GRB afterglows and
beaming-induced orphan afterglows.

\section {Photospheric emission}
In this study, we consider axisymmetric jet and assume that the observer is on the axis $\overline{OD}$ in
Fig. \ref{fig:cartoon}b. This is the extension of the 1-D formulation derived by Daigne \& Mochkovitch (2002).
For a baryon-rich ejecta in the stellar frame (i.e., burst source frame, and from now on all variables are defined in
this frame),
we assume the ejecta has been accelerated at a distance $r_{\rm acc}$ from the central engine.
The mass flux of the ejecta is written as $\dot{M}=\dot{E}/\Gamma c^2$,
where $\Gamma$ is the Lorentz factor of the ejecta and $\dot{E}$ is
the energy injection rate. The energy injection begins at $t_{\rm inj}$=0, and stops at
$t_{\rm inj}=t_{\rm w}$, i.e., the central engine activity lasts for a period of time $t_{\rm w}$.

The ejecta can be subdivided into a series of concentric layers,
where each layer has been injected at $r_{\rm acc}$ at a certain injection time $t_{\rm j}$.
Each ejecta layer becomes transparent when it has expanded to a distance
$r(t_{\rm inj}=t_{\rm j},t)$ at a specific time $t$, where
$r(t_{\rm inj}=t_{\rm j},t)=r_{\rm acc}+\beta c (t-t_{\rm j})$ and $\beta=\sqrt{1-1/\Gamma^{2}}$.
The photons emitted at $r(t_{\rm inj}=t_{\rm j},t)$
will escape from the ejecta at a time $t_{\rm esc}$ at a distance
$r_{\rm esc}(t_{\rm inj}=t_{\rm j},t)$.
Here $r_{\rm esc}$ and $t_{\rm esc}$ are defined where a photon
emitted at time t by the shell ejected at time $t_{\rm inj}=t_{\rm j}$ escapes the outflow, i.e. reaches
the first shell emitted at time $t_{\rm inj}=0$.
In other words, the optical depth from $r(t_{\rm inj}=t_{\rm j},t)$ to $r_{\rm esc}(t_{\rm inj}=t_{\rm j},t)$ is unity.

A geometric sketch illustrating the escape path of photons inside the ejecta is shown in Fig. 1b. A photon emitted
at point E (at a distance $r(t_{\rm inj}=t_{\rm j},t)$ and propagation angle $\varphi$) will escape from
the ejecta at point G, such that the optical depth from E to G is
\begin{equation}
  \tau(t_{\rm j},\varphi)=\int^{r_{\rm esc}(t_{\rm inj}=t_{\rm j},t)}_{r(t_{\rm inj}=t_{\rm j},t)}d\tau(r),
\end{equation}
where $d\tau(r)$ can be estimated as (Abramowicz et al. 1991; Daigne \& Mochkovitch 2002; Pe$'$er 2008)
\begin{equation}
  d\tau(r)=\frac{\kappa \dot{M} (1-\beta {\rm cos} \varphi)}{4\pi r^2}dr.
\end{equation}

We define the photospheric radius $R_{\rm PS}(t_{\rm j})=r(t_{\rm j},t)$ at which $\tau$ is equal to unity.
In light of Fig. \ref{fig:cartoon}b, we choose a cylindrical coordinate system (Pe$'$er 2008)
with the central point $O$ being the stellar center,
and the observer located along the +z-direction (defined by the direction of $\overline{OD}$).
Photons are emitted at a perpendicular distance $r_{\rm min}=r(t_{\rm j},t){\rm sin} \varphi$ from the z-axis and
a distance $z_{\rm min}=r(t_{\rm j},t){\rm cos} \varphi$ along the z-axis from point $O$.
The escape radius can be estimated from the triangle $OEG$, i.e.,
$r_{\rm esc}(t_{\rm inj}=t_{\rm j},t)=
r(t_{\rm j},t)+\beta c (t_{\rm j}-0)+\beta c(t_{\rm esc}-t)    
=\{r(t_{\rm j},t)^2+[c (t_{\rm esc}-t)]^2-
2 r(t_{\rm j},t) c (t_{\rm esc}-t) {\rm cos} (\pi-\varphi)\}^{\frac{1}{2}}$,

The integration for the optical depth can be conveniently rewritten in cylindrical coordinates as the following
\begin{equation}
  \tau(t_{\rm j},\varphi)=\int ^{z_{\rm max}}_{z_{\rm min}} \frac{\kappa \dot{M} (1-\beta {\rm cos} \varphi)}
  {4\pi r^2}\frac{dr}{dz} dz,
\end{equation}
where $r=\sqrt{z^2+r_{\rm min}^2}$, $dz/dr=z/\sqrt{z^2+r_{\rm min}^2}$ and
$z_{\rm max}=\sqrt{r_{\rm esc}^2-r_{\rm min}^2}$.
The photospheric radius ($R_{\rm PS}$), which depends on the propagation angle ($\varphi$),
can be found readily by defining $\tau=1$.

For a relativistic ejecta with Lorentz factor $\Gamma$, the arrival time of photons emitted at the photosphere
in the observer frame are delayed relative to that measured in the stellar frame
\begin{equation}
\label{equ:tobs}
    t_{\rm obs}=t-R_{\rm PS}{\rm cos}\varphi/c=t_{\rm j}+(1-\beta{\rm cos}\varphi)R_{\rm PS}/\beta c,
\end{equation}
i.e., the observer time is a function of injection time and propagation angle.
In this equation, we have neglected the effect of the acceleration radius ($r_{\rm acc}$) because
it is much smaller than the photosphere radius.

\begin{figure}
\hspace*{-0.8cm}\includegraphics[scale=0.8]{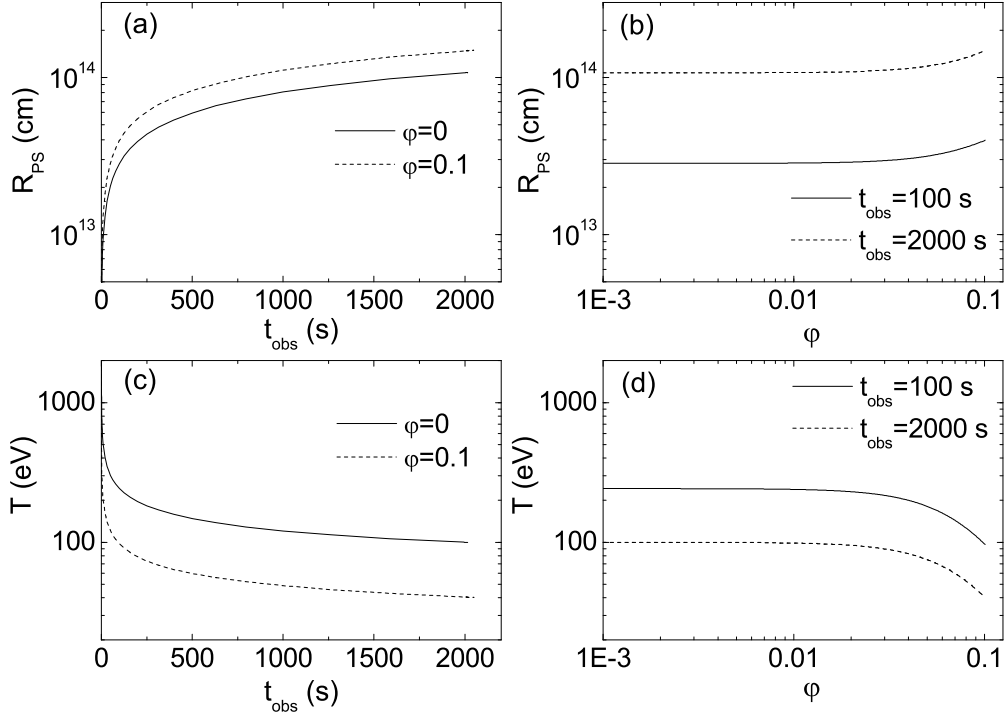}
\caption{\label{fig:4p}
Evolution of the photospheric radius and temperature with respect to the observer time and photon propagation angle.
The parameters are taken as $\Gamma=10$, $\dot{E}=10^{51}$erg/s and $t_{\rm w}=2000$s.}
(a) The photospheric radius vs. the observer time. The solid curve is plot for photons propagating
along the expansion direction of the ejecta ($\varphi=0$). The dashed curve corresponds to $\varphi=0.1$ rad.
(b) The photospheric radius vs. the photon propagation angle. The solid and dashed curves
correspond to the observer time of 100 s and 2000 s respectively.
(c) The photospheric temperature vs. observer time. The solid and dashed curves correspond to
photon propagation angles of 0 and 0.1 rad, respectively.
(d) The photospheric temperature vs. the photon propagation angle.
The solid and dashed curves are for $t_{\rm obs}$=100 s and 2000 s, respectively.
\end{figure}

The evolution of photospheric radius with propagation angle is shown in Fig. \ref{fig:4p}b.
The parameters are taken as $\Gamma=10$, $\dot{E}=10^{51}$erg/s and $t_{\rm w}=2000$s.
The solid curve shows its evolution at observer time $t_{\rm obs}=100$ s
and the dashed curve at $t_{\rm obs}=2000$ s. From this panel, we can see that in the
observer frame, the photospheric radius decreases with propagation angle.

We also show the evolution of photospheric radius with observer time in Fig. \ref{fig:4p}a. The solid, dashed curves
present the cases for $\varphi=0$ and $\varphi=0.1$ rad respectively. As is shown in this panel, the photospheric
radius increases with time, and the duration of the photospheric emission is prolonged at larger propagation angle.
The end points of the two curves indicate the observer time when the last layer of the ejecta becomes transparent.

According to the fireball model, the temperature of a layer at its photospheric radius is given by (Piran 1999)
\begin{equation}
  kT_{\rm PS}=\frac{D}{\Gamma}kT^{0}\big{(}\frac{R_{\rm PS}}{r_{\rm acc}}\big{)}^{-2/3},
\end{equation}
where $D=[\Gamma(1-\beta {\rm cos} \varphi)]^{-1}$ is the Doppler factor,
$r_{\rm acc}$ is the saturation radius and $T^0$ is its blackbody temperature.
In Fig. \ref{fig:4p}c and Fig. \ref{fig:4p}d, we show the evolution of $T_{\rm PS}$ with respect to the
propagation angle and observer time respectively. From the two panels, we find that
the photosperic temperature decreases with the propagation angle and time in the observer frame.

\begin{figure}
\hspace*{-0.8cm}\includegraphics[scale=0.8]{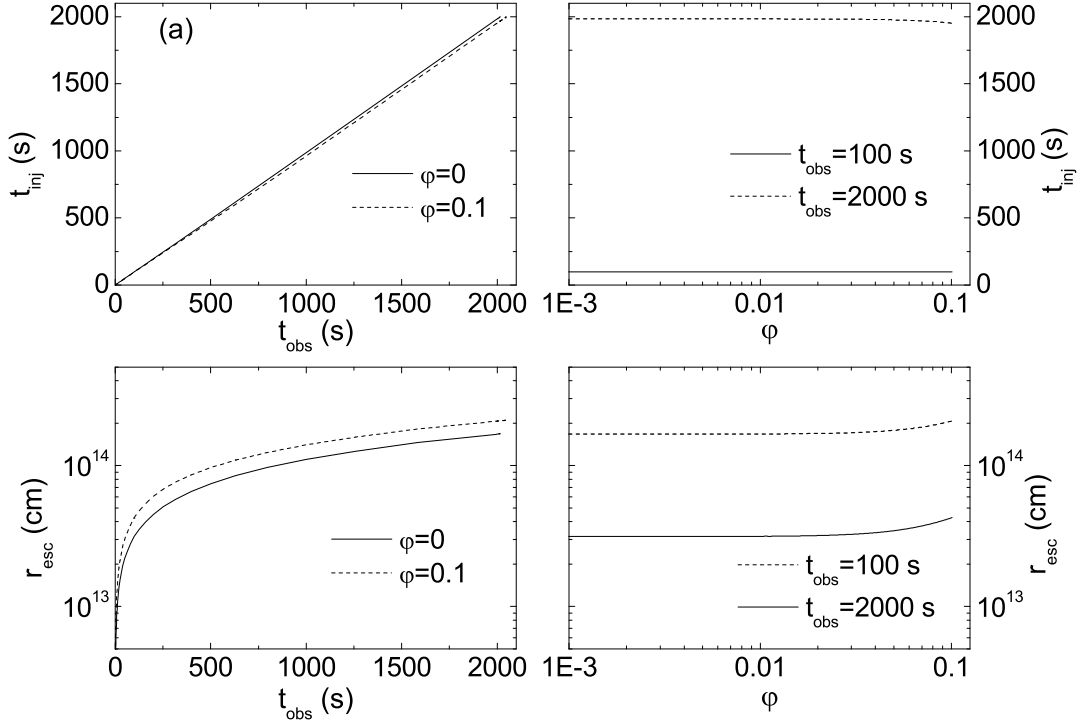}
\caption{\label{fig:4p2}
Evolution of the injection time and escaping radius with respect to the observer time and photon propagation angle.
The parameters are the same as those in Fig. \ref{fig:4p}.
(a) The injection time vs. the observer time. The solid curve is plot for photons propagating
along the expansion direction of the ejecta ($\varphi=0$). The dashed curve is the
evolution of the injection time for $\varphi=0.1$ rad.
(b) The injection time vs. the photon propagation angle. The solid and dashed curves are
observed at 100s and 2000s respectively.
(c) The escaping radius vs. the observer time. The solid and dashed curves are plot
with a propagation angle of 0 and 0.1 rad, respectively.
(d) The escaping radius vs. the photon propagation angle.
The solid and dashed curves are for $t_{\rm obs}$=100 s and 2000 s, respectively.}
\end{figure}

The evolution of the injection time with respect to the observer time
and the propagation angle is shown in Fig. \ref{fig:4p2}a and \ref{fig:4p2}b, respectively.
For the ``standard'' parameter set
($\Gamma=10$, $\dot{E}=10^{51}$ erg/s, $t_{\rm w}$=2000 s), $t_{\rm inj}(t_{\rm obs})$
is mildly smaller than $t_{\rm obs}$ for different $\varphi$, and
$t_{\rm inj}(\varphi)$ is almost independent of $\varphi$ for different $t_{\rm obs}$.
In Fig. \ref{fig:4p2}c and Fig. \ref{fig:4p2}d, we also show the evolution of the
escaping radius $r_{\rm esc}$ with respect to the observer time and the propagation angle,
respectively. The relations of $t_{\rm obs}$ and $\varphi$
with respect to $r_{\rm esc}$ are similar to that of the photospheric radius $R_{\rm PS}$.

As for a jet with half-opening angle $\theta=0.1$ rad, constant Lorentz factor $\Gamma=10$, and energy
injection from $t_{\rm inj}=0$ to $t_{\rm inj}=t_{\rm w}=2000$ s with energy injection rate per solid angle
$\dot{E}/4\pi=10^{51}/4\pi$ erg/s, we can estimate that $r_{\rm acc}\simeq 9\times 10^7$cm and
$kT^0\simeq0.41$ MeV for a fireball model
(Piran 1999; $\rm M\acute{e}sz\acute{a}ros~\&~Rees$ 2000; Daigne \& Mochkovitch 2002).
If the line-of-sight is along the jet central axis, we can find that the photospheric radius
is about $1.1\times10^{14}$ cm when the last layer becomes transparent, the observer's time
can be calculated from Eq. \ref{equ:tobs}, which is found to be about 2020 s and corresponds to the end point of
the solid curve in Fig. \ref{fig:4p}a.

\begin{figure}
\hspace*{-1.0cm}\includegraphics[scale=0.8]{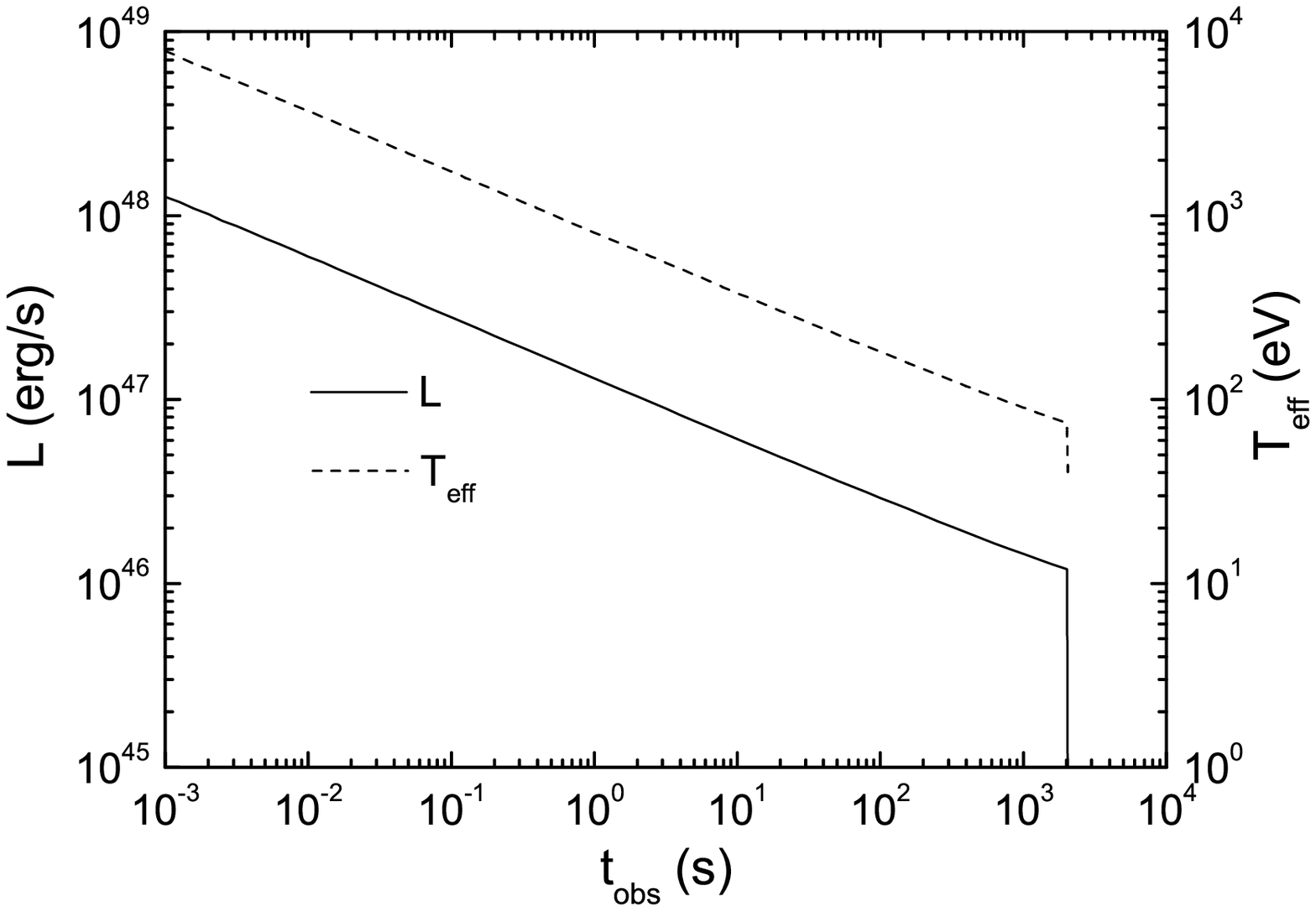}
\caption{\label{fig:LTt} Evolution of the photospheric luminosity (solid curve) and effective temperature (dashed curve)
with observer time for a jet with parameters of $\theta=0.1$ rad, $\Gamma=10$,
$\dot{E}=10^{51}$erg/s and $t_{\rm w}$=2000 s.
}
\end{figure}

The observed luminosity of photospheric emission can be determined by integrating over the surface of the photosphere
\begin{equation}
  L=\int_{0}^{\theta}\sigma T_{\rm PS}^4 dS {\rm cos} \vartheta 
\end{equation}
where $\vartheta$ is the angle between the tangential direction of the photosphere surface and the line-of-sight
when the propagation angle is $\varphi$.
 $dS {\rm cos} \vartheta=2\pi R_{\rm PS}(t_{\rm obs},\varphi){\rm sin} \varphi
[R_{\rm PS}(t_{\rm obs},\varphi+d\varphi){\rm sin} (\varphi+d\varphi)-R_{\rm PS}(t_{\rm obs},\varphi){\rm sin}\varphi]$
is the photospheric surface area from propagation angle $\varphi$ to $\varphi+d\varphi$.
The evolution of the photospheric luminosity with observer time is shown by the solid curve in Fig. \ref{fig:LTt}.
There is a break in the light curve at about $t_{\rm obs}\simeq$ 2020 s, which is attributed to the stop of energy
injection by the central engine and when the last layer of the ejecta became transparent as the photons propagate along the line-of-sight.
Afterwards, only photospheric emission at high latitude (large propagation angles) contributes to the observed luminosity. The photospheric emission ceases when the last layer with propagation angle $\varphi=\theta=0.1$ become transparent, which is about 
2050 s in the observer frame.

We can also define an effective temperature for the photosphere
\begin{equation}
  T_{\rm eff}=\frac{\int_{0}^{\theta}T dL}{\int_{0}^{\theta}dL}.
\end{equation}
This effective temperature is shown as a dashed curve in Fig. \ref{fig:LTt}. We can find that
the photospheric emission of a failed GRB is presented as a short soft X-ray burst and then becomes a
UV burst which lasts for about several thousand seconds.

\begin{figure}
\hspace*{-0.8cm}\includegraphics[scale=0.8]{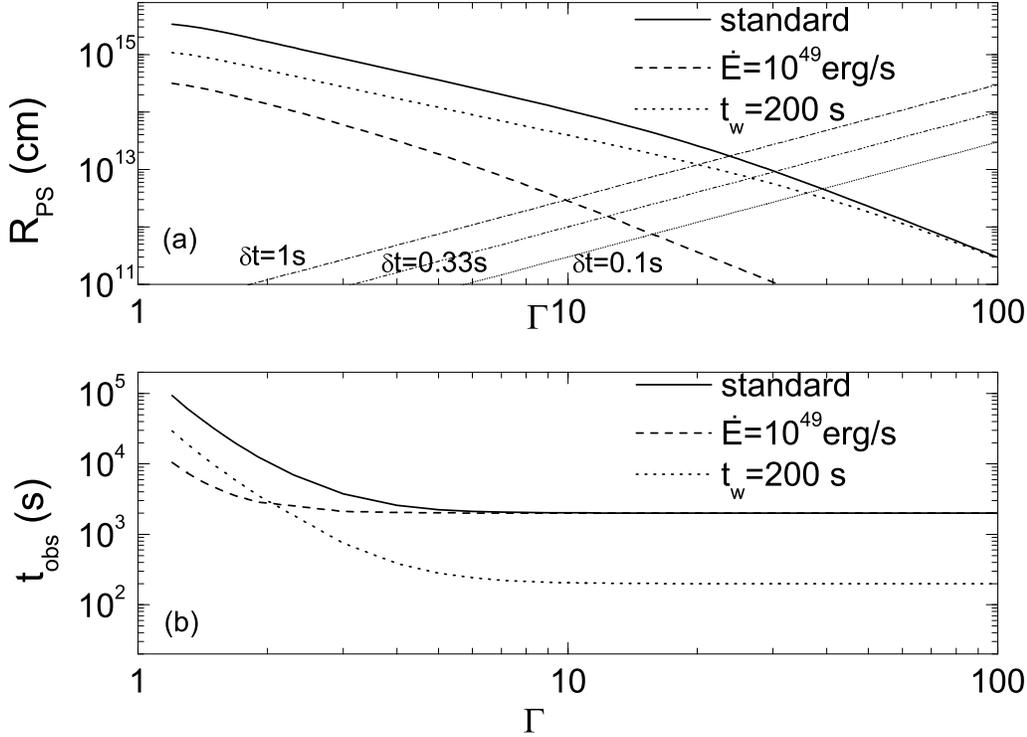}
\caption{\label{fig:Rtg} Parameter dependence of the photospheric emission. 
All curves are obtained when 
the last layer of the ejecta became transparent along the line-of-sight,
i.e., $\varphi=0$ and $t_{\rm j}=t_{\rm w}$.
(a) Parameter dependence of the photospheric radius. The solid curve is derived using the standard
parameters, i.e., $\dot{E}=10^{51}$erg/s and $t_{\rm w}$=2000~s; the dashed curve is for
$\dot{E}=10^{49}$erg/s and $t_{\rm w}$=2000~s; and the dotted curve is for
$\dot{E}=10^{51}$erg/s and $t_{\rm w}$=200~s.
The evolution of the internal shock radii $R_{\rm IS}$ for a variability timescale of
$\delta t$=1 s, 0.33 s and 0.1 s are shown and marked correspondingly.
(b) Parameter dependence of the observer time. The identities of the curves are the same as in (a).
}
\end{figure}

\begin{figure}
\hspace*{-0.8cm}\includegraphics[scale=0.8]{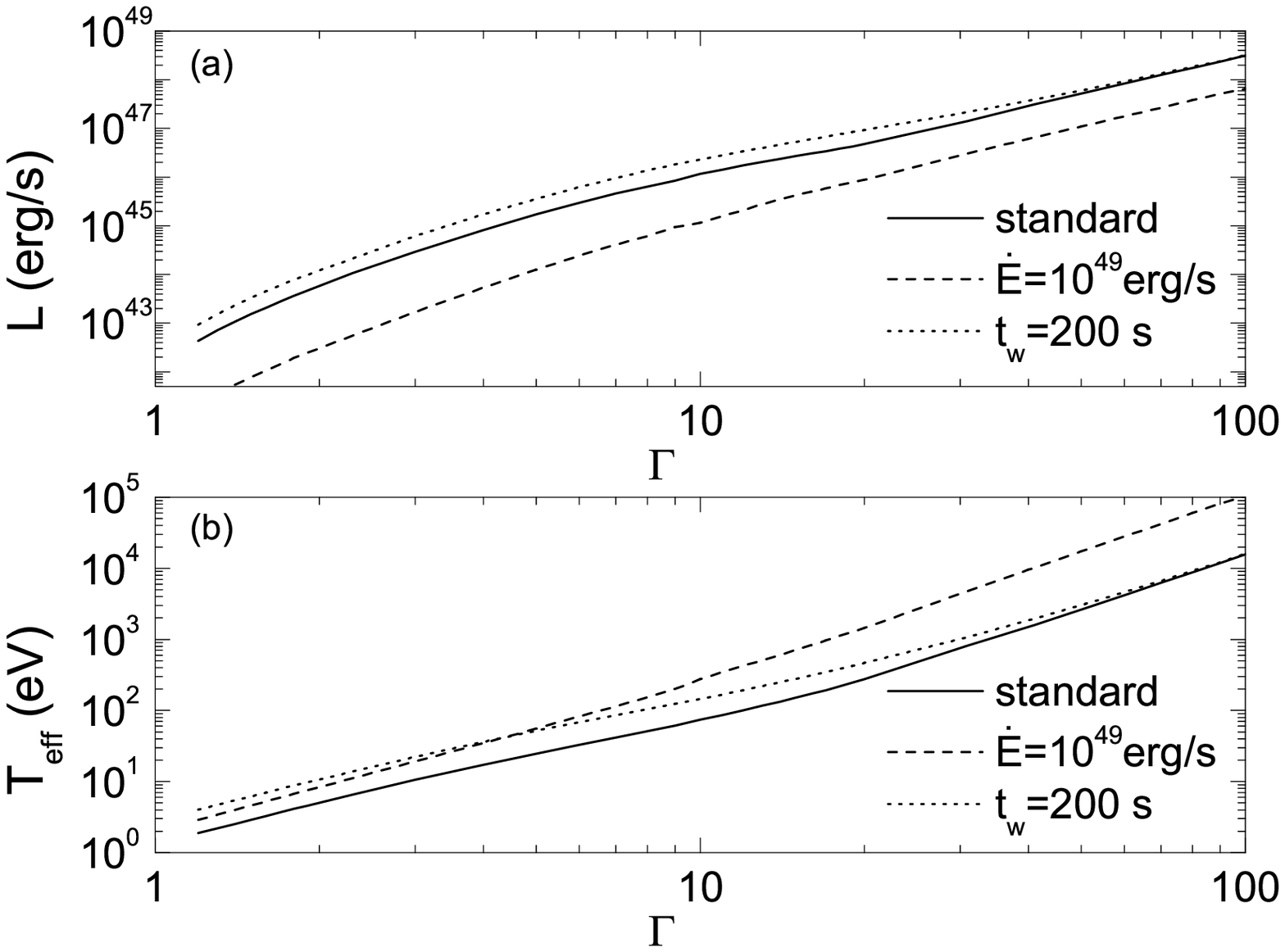}
\caption{\label{fig:LTg} Parameter dependence of the photospheric luminosity
(a) and the effective temperature
(b) in the Lorentz factor space. All curves are obtained when
the last layer of ejecta became transparent along the line-of-sight,
i.e., $\varphi=0$ and $t_{\rm j}=t_{\rm w}$.
The identities of the curves are the same as in Figure \ref{fig:Rtg}.
}
\end{figure}

We also investigated the parameter effect on the photospheric emission,
which are shown in Fig. \ref{fig:Rtg} and Fig. \ref{fig:LTg}.
All the curves are derived
when the last layer of ejecta along the line-of-sight became transparent
, i.e., $\varphi=0$ and $t_{\rm j}=t_{\rm w}$.
As is shown in Fig. \ref{fig:Rtg}a, the photospheric radii are decreasing with
the increase of Lorentz factor for different sets of parameters.
The solid curve corresponds to the standard parameters ($\dot{E}=10^{51}$erg/s, $t_{\rm w}=2000$ s),
while the parameters for the dashed curve and the dotted curve are $\dot{E}=10^{49}$erg/s, $t_{\rm w}=2000$ s and
$\dot{E}=10^{51}$erg/s, $t_{\rm w}=200$ s, respectively. A lower energy injection rate and shorter injection
time will decease the radius of the photosphere.
Fig. \ref{fig:Rtg}b shows the evolution of the observer time with Lorentz factor.
The parameters for each curve are the same as Fig. \ref{fig:Rtg}a.
This time period can be interpreted as the duration of the photospheric emission.
From this panel, we can find that the duration of photospheric emission decreases with an increase of Lorentz factor.
Lower energy injection rate and shorter injection time will decease the duration
of the photospheric emission.

In Fig. \ref{fig:LTg}, we show the parameter effect on the photospheric luminosity and effective temprature.
The parameters of each curve are the same as Fig. \ref{fig:Rtg}. From Fig. \ref{fig:LTg}a, we can find that the luminosity
of the photospheric emission are low in both high and low Lorentz factor. Lower energy injection
rate results in lower luminosity. As is shown in Fig. \ref{fig:LTg}b, the effective temperatures are
increasing with the increase of Lorentz factor. Lower energy injection rate and shorter injection
time will decease the radii of the photosphere and hence results in a higher effective temperature.

As for a jet with half-opening angle of about 0.1 rad,
Lorentz factor $\Gamma=2-20$, energy injection rate $\dot{E}=10^{49-51}$erg/s and injection time $t_{\rm w}$=200-2000s,
from Fig. \ref{fig:Rtg} and \ref{fig:LTg}, we can conclude that the prompt emission for a failed GRB
is thermal soft X-ray or UV photospheric emission, there will be no significant non-thermal gamma-ray emission.
The photospheric luminosity is about $10^{46}$erg/s and last for about one thousand seconds.
Note that the photospheric luminosity is far lower than the energy injection power,
most of of the energy is re-converted into the ejecta's kinetic energy.
From Fig. \ref{fig:Rtg}a, we find that the radius of the photosphere is about $10^{14}$ cm, which is larger
then the prediction for the internal shock's radius, i.e.,
$R_{\rm IS}\simeq\Gamma^2 c \delta t\simeq 10^{12}$ cm (M\'{e}sz\'{a}ros 2006),
where $\delta t\sim0.33$ s here is the variability timescale of the prompt emission.
The evolution of $R_{\rm IS}$ with respect to $\Gamma$ for $\delta t$=1 s, 0.33 s and 0.1 s
are shown and marked in Fig. \ref{fig:Rtg}a correspondingly.
This radius is consistent with Fig. \ref{fig:cartoon}a.

This thermal radiation will be in the UV or soft
X-ray band. The lower band of Swift-XRT ($0.2-10$ keV) may cover this energy range and it is
sensitive enough to detect such a photospheric emission component (Gehrels et al. 2004). MAXI-SSC
monitors all-sky in the energy range of $0.5-10$ keV and also has a chance to detect such
events (Matsuoka et al. 1997). Future UV satellites may be also have the capability to detect these events,
such as TAUVEX (wavelength range 120nm-350nm) (Safonova et al. 2008).

\section{Afterglow emission}
As the outflow expands outward, it will collide with the surrounding medium and
afterglow will be produced. The dynamical evolution of a
relativistic jet in interstellar
medium has been studied by Huang et al. (1999). Their codes
can be used in both ultra-relativistic and non-relativistic phases.

In our model, we consider a jet with the bulk Lorentz factor $\Gamma=10$,
the half-opening angle $\theta=0.1$ and the isotropic energy
$E=10^{50}$ erg. The jet expands laterally at the
comoving sound speed and collides with a medium whose number density is
$n_{\rm ISM}=1$ cm$^{-3}$.
We also assume typical values for some other parameters of the jet, i.e., the electron energy fraction
$\epsilon_{e}=0.1$, the magnetic energy fraction $\epsilon_{B}=0.01$ and the
power-law index of the energy distribution function of electrons $p=2.5$.
Multiband afterglow emission is expected from synchrotron radiation of relativistic
electrons. Using this exquisite model, we numerically calculated the afterglow light
curves and spectra with line of sight parallel to the jet axis. We assume a
redshift $z=1$ and a standard cosmology with $\Omega_{M} = 0.27$, $\Omega_{\Lambda}=0.73$
and with the Hubble constant of $H_{0}=71$ km s$^{-1}$ Mpc$^{-1}$.

\begin{figure}
\hspace*{-1.3cm}\includegraphics[scale=0.8]{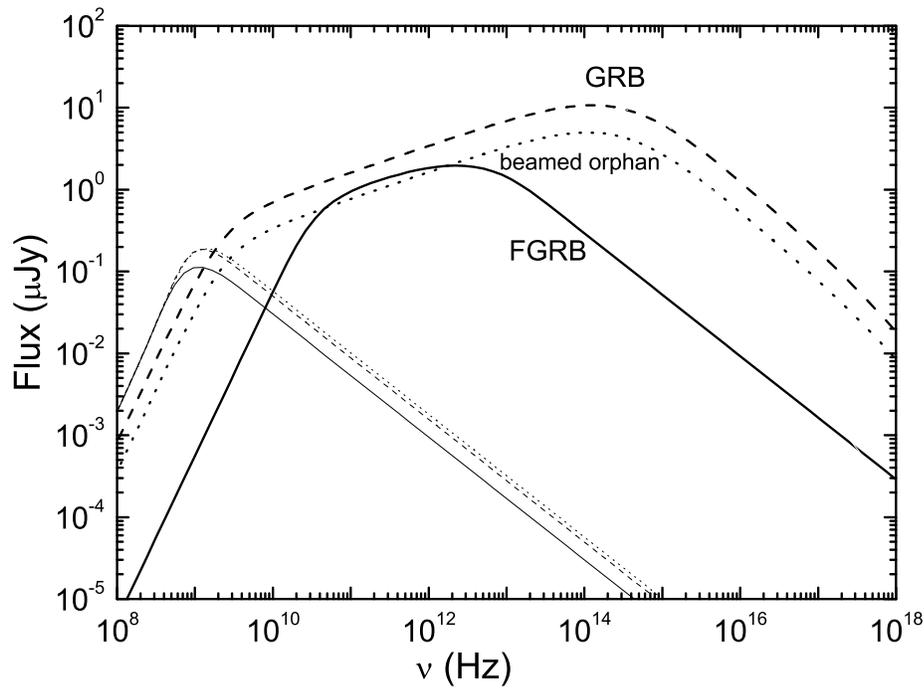}
\caption{\label{fig:spectra} Evolution of the afterglow spectra for the three types of GRBs.
The solid, dashed and dotted curves are spectra of a failed GRB afterglow,
an ordinary GRB afterglow and a beaming-induced orphan afterglow respectively.
The thick and thin curves are the spectra observed at $10^{3}$ s
and $10^{6}$ s respectively.
}
\end{figure}

Our results for the afterglow spectra of failed GRBs are shown in Fig. \ref{fig:spectra}.
The thick and thin solid curves are the spectra observed at $10^{3}$ s
and $10^{6}$ s respectively. In this figure, we can find that
the spectrum becomes softer with the elapse of the observational time. At the
early stage ($10^{3}$ s), the peak flux appears at about
$5\times10^{12}$ Hz, i.e.,
in the IR band. Both the peak flux and peak frequency decrease with time. At
late time ($10^{6}$ s), the peak flux is more than one magnitude less than
that in the early stage. The peak frequency decreases to about
$10^{9}$ Hz at late stage.
It is in the radio band and more than three magnitudes less than the peak frequency at the early
stage.

For comparison, we also show the afterglow spectra
of an ordinary GRB in Fig. \ref{fig:spectra}. Here we choose the same parameters as the failed GRB
except for a much larger bulk Lorentz factor ($\Gamma=300$). The dashed curves show
the afterglow spectra of this GRB with observing angle $\theta_{\rm obs}=0$ (the line of
sight is parallel to the jet axis). As is shown in Fig. \ref{fig:spectra},
the spectra of the failed GRB and the ordinary GRB are similar at the late stage
(thin solid and thin dashed curves) because their energies
and Lorentz factors are both similar at this moment.
But at early stage, they
are very different (thick solid and thick dashed curves)
due to their very different initial Lorentz factors and the
corresponding minimum Lorentz factors of electrons (Sari et al. 1998).
The peak frequency of the GRB afterglow
is much larger than that of the failed GRB afterglow.

In Fig. \ref{fig:spectra}, we also show the spectra of
a beaming-induced orphan afterglow
(afterglow from an ordinary highly collimated
GRB outflow, but with the observing angle larger than the jet half-opening angle so that
no prompt gamma-rays can be observed in the main burst phase,
Rhoads 1997; Huang et al. 2002). Here we assume the
same parameters as the ordinary GRB except $\theta_{\rm obs}=0.125$.
The early and late spectra of this orphan
afterglow are shown in Fig. \ref{fig:spectra} with thick dotted curve and thin dotted curve.
From this figure, we can find the spectra of beaming-induced orphan afterglow are
similar to that of the ordinary GRB. Although it is hard to distinguish
a beaming-induced
orphan afterglow from a failed GRB afterglow through their afterglow light curves
(Huang et al. 2002), they can be potentially distinguished from their
spectra at the early stages.
Their spectra of early afterglows are very different: the peak frequency of
a failed GRB afterglow is far lower than that of
a beaming-induced orphan afterglow.
Another way to distinguish them is through
their early light curves. Early afterglow of
a beaming-induced orphan afterglow will show
a rebrightening while failed GRB will not (Huang et al. 1999, 2002; Xu \& Huang 2010).

\section {Conclusion and Discussions}

The analysis in this paper shows that the emission of ejecta with low Lorentz factors is
very different from that expected from ejecta with high Lorentz factors.
Prompt emission of a GRB is non-thermal and bright in the gamma-ray band.
For a failed GRB, however, the emission originates from the photosphere with a thermal spectrum, and is bright
in the UV or soft X-ray band instead of gamma-rays. This photospheric emission lasts for about a thousand seconds
with a luminosity about several times $10^{46}$ erg/s.

Since the photospheric emission manifests as a UV or soft X-ray transient,
it can be detected by some current and future satellites, such as Swift-XRT, MAXI-SSC
and TAUVEX etc. On 2008 January 9, Swift-XRT discovered a peculiar
X-ray transient 080109 in NGC 2770 (Berger \& Soderberg 2008; Page et al. 2008).
No gamma-ray emission was detected.
This X-ray transient reached its peak at about $60s$ and lasted for about $600$ s.
Its spectrum can be fitted with an absorbed double blackbody model with temperatures about
$0.36$ keV and $1.24$ keV respectively (Li 2008). This transient may be a candidate of photospheric
emission from a failed GRB. Meanwhile, some unidentified X-ray transients have been detected by MAXI
during its one-year monitoring (Nakajima et al. 2009; Suzuki et al. 2010).
These transients generally showed an absorbed blackbody
spectrum and lasted for tens of seconds. It is possible that some of them
are photospheric emission from failed GRBs.

If we extend the injection time to about $10^{5}$ s and the jet half-opening to about 0.4 rad in our model,
we find that the photospheric radius is about $10^{15}$ cm and the effective temperature is deceased to lower
than 1 eV, i.e., there will be an \textit{optical burst}. This kind of optical burst will last for about several thousand
seconds with a luminosity about $10^{42}$ erg/s, which may be detected by the Hyper-Suprime Camera of the $Subaru$ telescope in the future.

In this work, we have assumed that the prompt emission is thermal radiation coming from the
photosphere where the optical depth is unity. Due to the low density of 
GRB jets,
however, it has been pointed out that the last-scattering positions of the observed photons
may not simply coincide with the photosphere, but instead possess a finite distribution around
it (e.g. Pe$'$er et al. 2006; Pe$'$er 2008; Beloborodov 2010; Pe$'$er \& Ryde 2011). This
stochastic effect can lead to differentiation of the observed spectrum from a thermal one
of purely photospheric origin.
Such mechanism can work even in failed GRBs, and it is our future work to study how the
spectrum will be reshaped using Monte-Carlo calculations.
We are planning to investigate this effect in the context of
failed GRBs as a next step of our study.

From the comparison of afterglow emissions from failed and ordinary GRBs,
while we find it challenging to distinguish them at their late stage of evolution, their spectra at the early stage
are profoundly different. We conclude that it is possible to identify
failed GRBs by observing their afterglow emission in the early stage.
The typical frequency at peak flux in the afterglow phase for failed GRBs is much lower
than that for ordinary GRBs
(or beaming-induced orphan GRBs). We can thus define a hardness
ratio, for instance, as the flux contrast between $10^{12}$ Hz and $10^{14}$ Hz at an observed time of $1000$
s, i.e.,
$f_{\rm 1ks}\equiv F_{10^{12}{\rm Hz}}/F_{10^{14}{\rm Hz}}$. If $f_{\rm 1ks}>1$, then it is quite likely that
the emission is coming from
a failed GRB. If $f_{\rm 1ks}<1$, then it would be more likely to come from an ordinary GRB afterglow
or a beaming-induced orphan afterglow. In addition, at the early afterglow stage, a rebrightening
phase will be present in the case of a beaming-induced
orphan GRB, while it is not expected for ordinary or failed GRBs.
Therefore, the afterglows of failed GRBs can be distinguished from both ordinary GRB afterglows
and beaming-induced orphan afterglows through observations at the early stages.

\acknowledgements
We thank the anonymous referee for many of the useful suggestions and comments.
We also would like to thank P. M\'{e}sz\'{a}ros, T. Piran and J. Aoi for helpful discussions.
This work was supported by the National Natural Science Foundation of China
(Grant No. 11033002), the National Basic Research Program of China (973 Program, Grant
No. 2009CB824800) and the Grant-in-Aid for the 'Global COE Bilateral
International Exchange Program' of Japan, Grant-in-Aid for Scientific
Research on Priority Areas No. 19047004 and Scientific Research on Innovative
Areas No. 21105509 by Ministry of Education, Culture, Sports,
Science and Technology (MEXT), Grant-in-Aid for Scientific Research (S)
No. 19104006 and Scientific Research (C) No. 21540404
by Japan Society for the Promotion of Science (JSPS).

\end{document}